\documentclass[12pt,preprint]{aastex} 
\usepackage{emulateapj5}
\usepackage{xspace}

\newcommand{\CGRO}{{\it CGRO}\xspace}
\newcommand{\RXTE}{{\it RXTE}\xspace}
\newcommand{\BATSE}{{BATSE}\xspace}
\newcommand{\ASM}{{ASM}\xspace}

\newcommand{\sS}[1]{\mbox{$\rm{}^{#1}$}}
\newcommand{\Ss}[1]{\mbox{$\rm{}_{#1}$}}

\renewcommand{\arcdeg}{\mbox{$^\circ$}\xspace}

\newcommand{\Ms}{\mbox{$M_{\odot}$}\xspace}
\newcommand{\Rs}{\mbox{$R_{\odot}$}\xspace}

\newcommand{\redchisq}{\mbox{$\chi^2_{\mbox{\scriptsize{red}}}$}\xspace}

\def\grs1758{\mbox{GRS~1758--258}\xspace}
\def\1e1740{\mbox{1E~1740.7--2942}\xspace}
\def\4u1700{\mbox{4U~1700--377}\xspace}

\begin{document}

\title{Instability of the 13.8 day period in X-ray emission from 4U~1700--377}

\shortauthors{Hong \& Hailey}
\slugcomment{for ApJ}

\author{
JaeSub Hong\altaffilmark{1} and
Charles J. Hailey\altaffilmark{2} 
}
\email{jaesub@head-cfa.cfa.harvard.edu}
\altaffiltext{1}{Harvard-Smithsonian Center for Astrophysics, 60 Garden St.,
Cambridge, MA 02138}
\altaffiltext{2}{Columbia Astrophysics Laboratory, 538 W. 120th St.,
New York, NY 10027}

\begin{abstract} 

We present a new result on long-term periodicity searches of the X-ray
emission from \4u1700 using \RXTE \ASM and \CGRO \BATSE data.  The hard
X-ray data (20 -- 200 keV) from early \BATSE observations (1780 days
from JD 2448370, before \RXTE observations started) show evidence of a
13.8 day periodicity. The long term periodicity became substantially
less prominent in the data from later \BATSE observations and in the
soft X-ray data from the \ASM (2 -- 10 keV) observations.  We
demonstrate that disk precession models can explain the 13.8 day period
and its instability in the X-ray emission from \4u1700.

\end{abstract}

\keywords{X-ray binary, long term period, disk accretion}

%\psdraft
\section{Introduction}

First discovered by \citet{Jones73}, \4u1700 is a 3.4 day X-ray
eclipsing binary system. Its optical counterpart, HD 153919, is a
bright early-type star. The system has been extensively studied at both
optical and X-ray wavelengths. While it is considered to be a neutron
star binary system there has been no confirmed detection of
pulsation.  

\citet{Konig97} reported a possible 13.8 day periodicity in the
X-ray emission from \4u1700.  Their results are based on the long term
observations by the All-Sky Monitor (\ASM, 2 -- 10 keV) \citep{Levin96}
on the Rossi X-ray Timing Explorer (\RXTE) and the Burst And Transient
Source Experiment (\BATSE, 20 -- 200 keV) \citep{Fishman89, Harmon02} on
the Compton Gamma Ray Observatory (\CGRO).  More data has been collected
since their report.  Here we present the results of long-term period
searches using all the data available as of this writing.

\section{Data selection and analysis}

We employed All-Sky monitor X-ray light curves from \BATSE on
\CGRO satellite and \ASM on the \RXTE satellite. All data
presented here are publicly available by an on-line service from the
Laboratory for High Energy Astrophysics (LHEA) at
NASA/GSFC\footnote{\noindent {\tt http://heasarc.gsfc.nasa.gov/} (HEASARC),  \\
{\tt http://cossc.gsfc.nasa.gov/archive/} (BATSE), \\
{\tt http://heasarc.gsfc.nasa.gov/docs/xte/xhp\_achive.html} ({\it RXTE})}.

\begin{figure*} 
\begin{center}
\epsscale{1.4}
%\epsscale{0.80}
\plotone{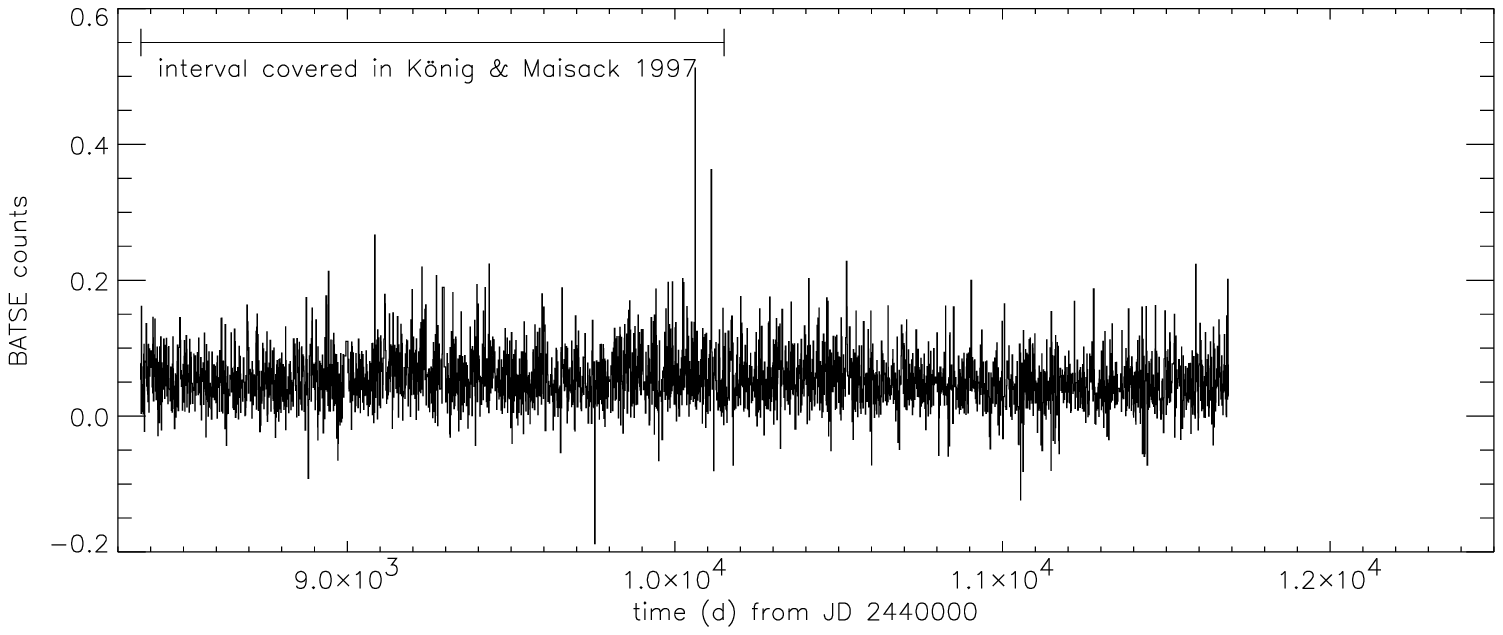}
\plotone{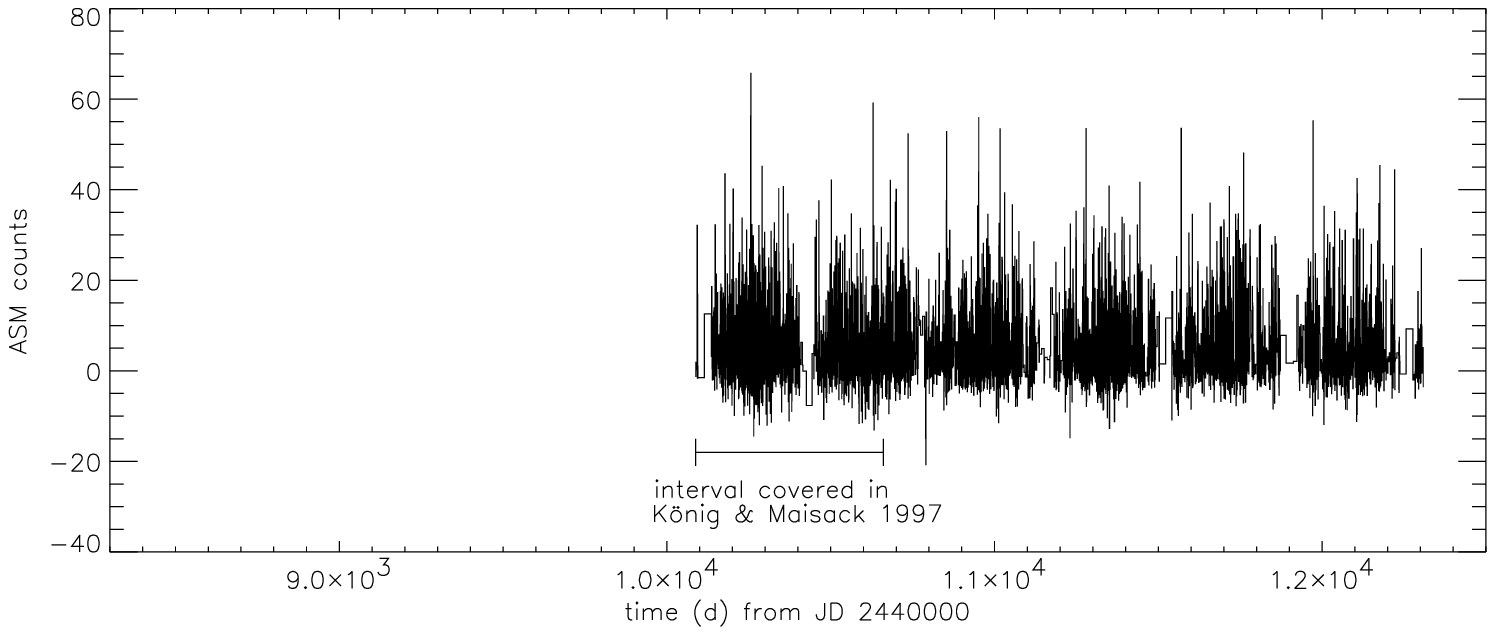}
\end{center}
\caption{X-ray light curve from \4u1700 : \CGRO \BATSE light curve by
occultation technique (top : 20 -- 200 keV)
and \RXTE \ASM light curve (bottom : 2 -- 10 keV). 
The intervals covered in \citet{Konig97} are marked by a horizontal bar.}
\label{fig:lc}
\end{figure*}

Using Earth occultations \BATSE provides long term flux histories for
bright X-ray sources in the energy range from 20 to 300 keV. In this paper, we
used 3321 days (from JD 2448370) of daily-averaged, long-term flux from
\4u1700 given by the \BATSE occultation technique (Figure~\ref{fig:lc}). 
The \ASM on the \RXTE satellite usually dwells on a bright source
a few times a day and it covers an energy range from 2 to 10 keV
(Figure~\ref{fig:lc}).  We used 2221 days (from JD 2450088) of \ASM
X-ray time series data.  No additional cut was applied to the data 
for analysis.

For the period search, \citet{Konig97} used the epoch-folding technique,
which is based on \citet{Leahy83}. The epoch-folding technique in
\citet{Leahy83} does not take into account phase selection effects.
Here we used the phase-averaged epoch-folding technique \citep{Collura87}.  The
phase-averaged epoch-folding technique is less vulnerable to false alarms
due to random fluctuations than a single-phase epoch-folding technique.
In practice, the outcome of the phase-averaged epoch-folding technique
is similar to that of the epoch-folding technique with an optimally
chosen bin size \citep{Collura87}.

\begin{table*}
\caption{Estimate of the orbital period $P_0$ and its width $\Delta P_0$
in the epoch-folding diagram}
\label{tab:orbital}
\begin{tabular}{r|ccr}
\hline \hline
Data set	& $P_0$ estimate\sS{a}	& $\Delta P_0$ estimate\sS{a}/theory\sS{b}& \redchisq\sS{c} \\
	days	& day 			& $\times 10^{-3}$ day 	  		&  	\\
\hline
\BATSE\   3321 	& 3.41155(10)		& 3.11(5)/3.50	  			& 58.2	\\
\ASM\ 2221	& 3.41161(05)		& 5.41(1)/5.24	  			& 366.4	\\
\hline
\end{tabular}
\\
\sS{a}$P$ and $\Delta P$ is calculated for the sinusoidal signal \citep{Leahy87}. \\
\sS{b}$\Delta P$ = $P^2$/$T$, where $T$ is the time span of the data set. \\
\sS{b}The maximum value of the peak\\
\end{table*}

\begin{figure*} 
\begin{center}
\epsscale{1.4}
%\epsscale{0.80}
\plotone{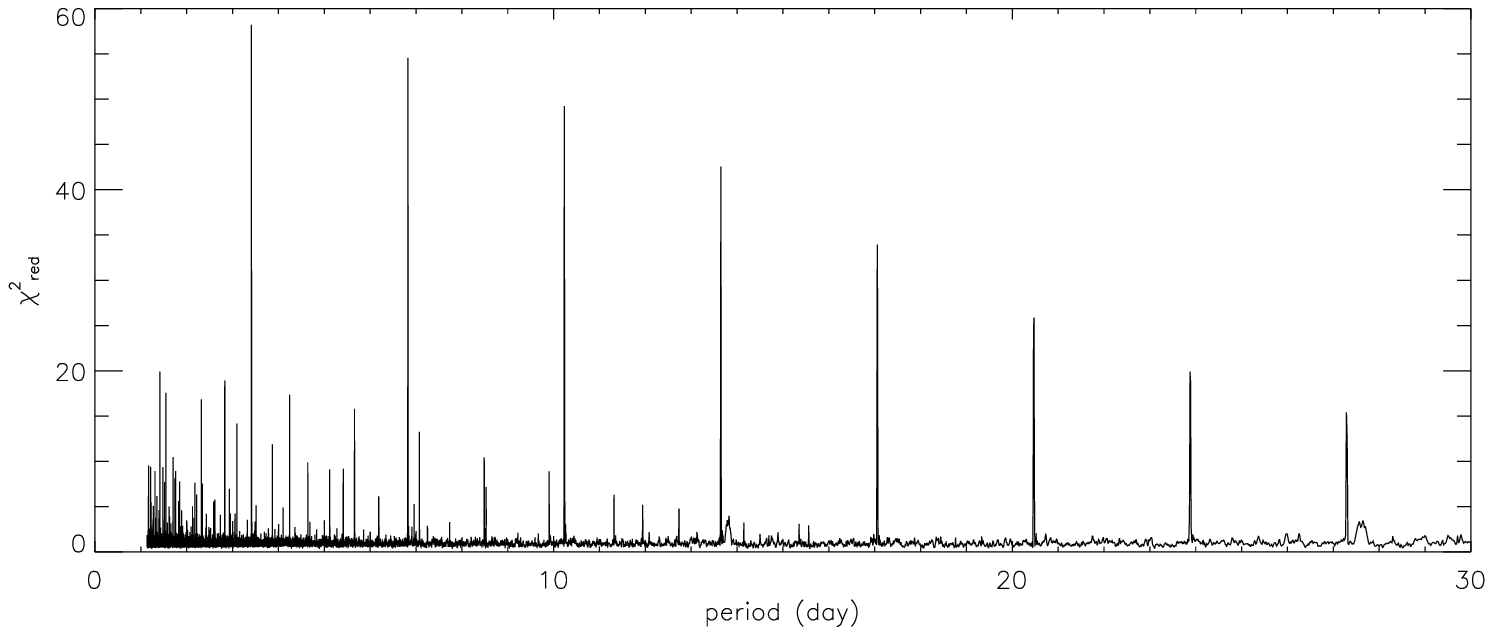}
%\epsscale{0.80}
\plotone{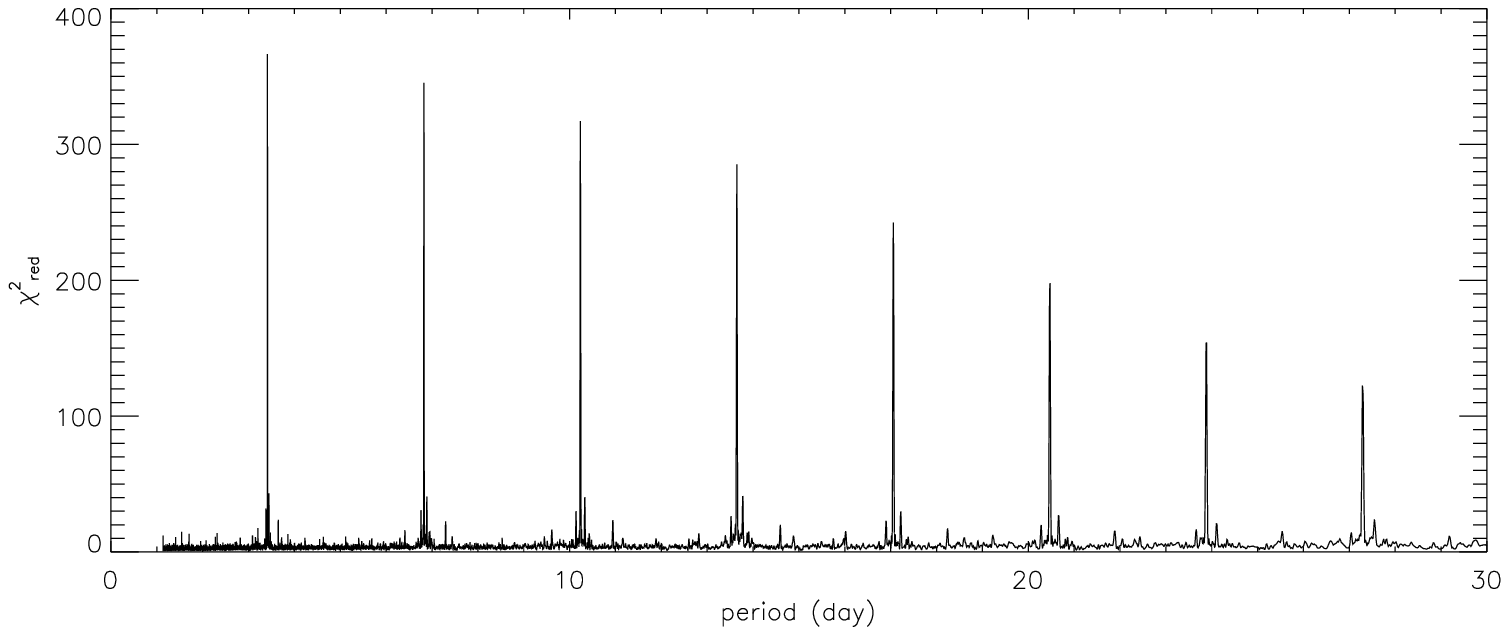}
\end{center}
\caption{The phase-averaged epoch folding results of the light curve from 
the \CGRO \BATSE data (top: 3321 days from JD 2448370.0) and the \RXTE \ASM data
(bottom: 2221 days from JD 2450088.6) (15 phase bins)}
\label{fig:efresults}
\end{figure*}

Figure~\ref{fig:efresults} shows the phase-averaged epoch folding
results for the light curves of \4u1700 in Figure~\ref{fig:lc} (15
phase bins). In Table~\ref{tab:orbital}, we estimated the orbital period
($P_0$), assuming the sinusoidal signal of light curves \citep{Leahy87}.
It should be noted that the true orbital
period varies as $\dot{P}/P = -3.3 \times 10^{-6}$~/yr
\citep{Rubin96}.

\begin{figure*} 
\begin{center}
\epsscale{1.4}
%\epsscale{0.80}
\plotone{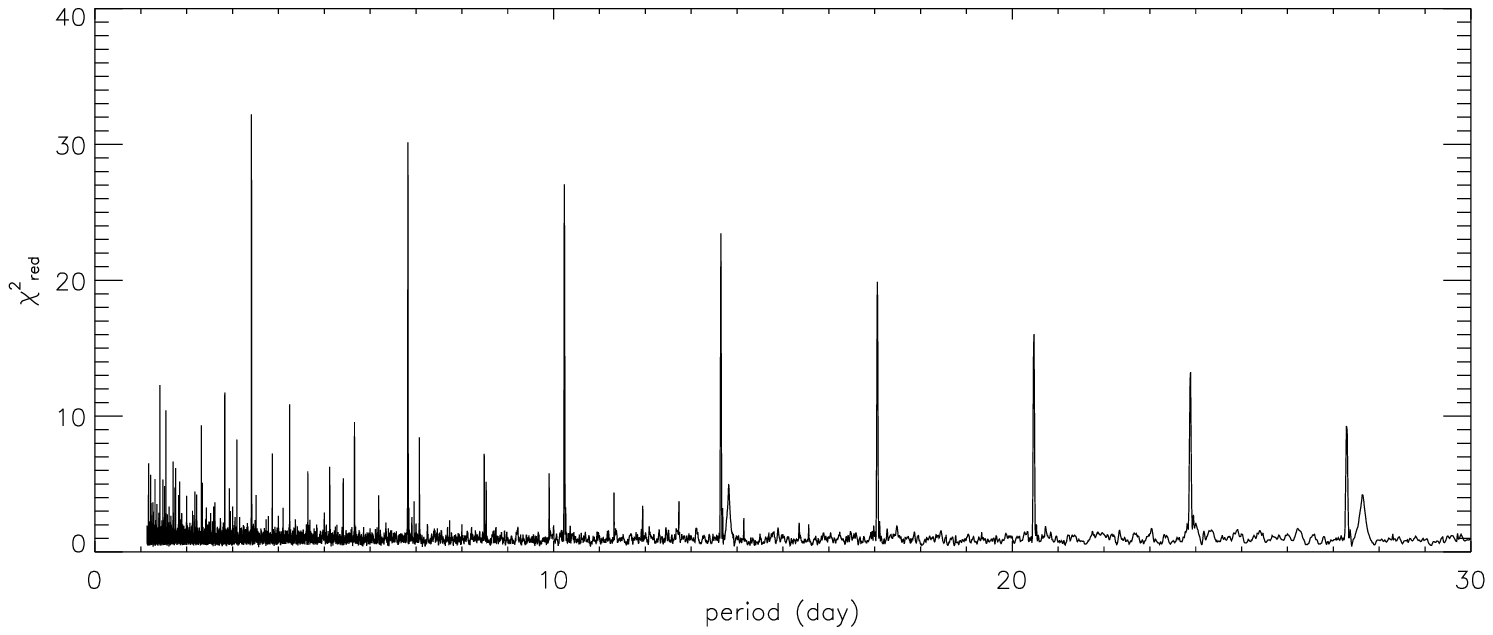}
\end{center}
\caption{The phase-averaged epoch folding results 
for the first 1780 days (from JD 2448370) of the \BATSE data: 
compare this plot with the figure~1b in \citet{Konig97}, where \redchisq at
$P_i$ is not always larger than that at $P_{i+1}$ due to the noise
fluctuations. One can also notice a broad peak around 27.6 days, which is
the second harmonics of the 13.8 day period. }
\label{fig:efphave}
\end{figure*}

In comparison with the previous report \citep{Konig97}, the
phase-averaged epoch folding on the 1780 days of \BATSE data
(Figure~\ref{fig:efphave}) produces roughly the same level of the noise
fluctuations as in Figure~1b of \citet{Konig97}, but \redchisq at the
multiples of the orbital period is about twice as large as those of
\citet{Konig97}. This is due to the superior performance of
phase-averaging in the epoch-folding technique.  In the case of the
first 573 days of the \ASM data, the noise fluctuation of \redchisq
by the phase-averaged epoch folding is dramatically smaller than that
of the previous reports (Figure~3 in \citet{Konig97}).  In
Figure~\ref{fig:efphave}, the second harmonics of the 13.8 day period is
also prominent as a broad peak around 27.6 days.

\begin{figure*} 
\begin{center}
%\begin{tabular}{c}
%\epsfig{figure=f4a.eps,width=6.0cm,clip=} 
%\epsfig{figure=f4b.eps,width=6.0cm,clip=} \\
%\epsfig{figure=f4c.eps,width=6.0cm,clip=}  
%\epsfig{figure=f4d.eps,width=6.0cm,clip=} \\
%\end{tabular}
\epsscale{0.7}
%\epsscale{0.40}
\plotone{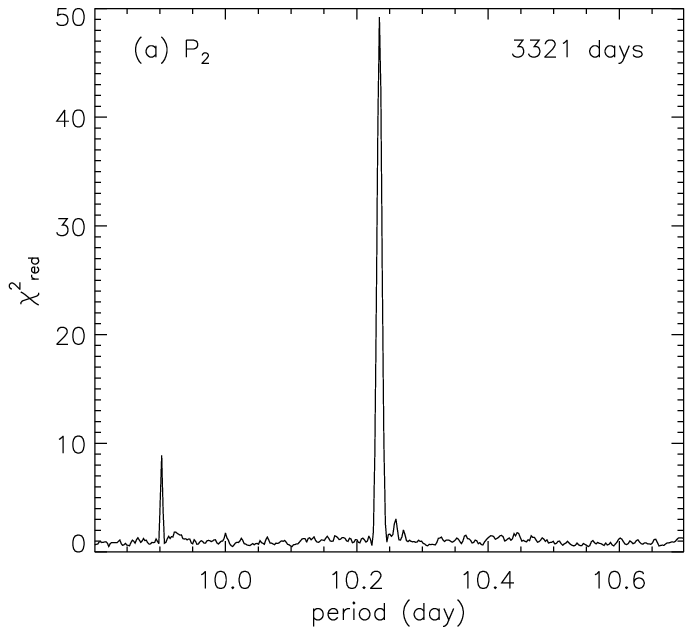}
\plotone{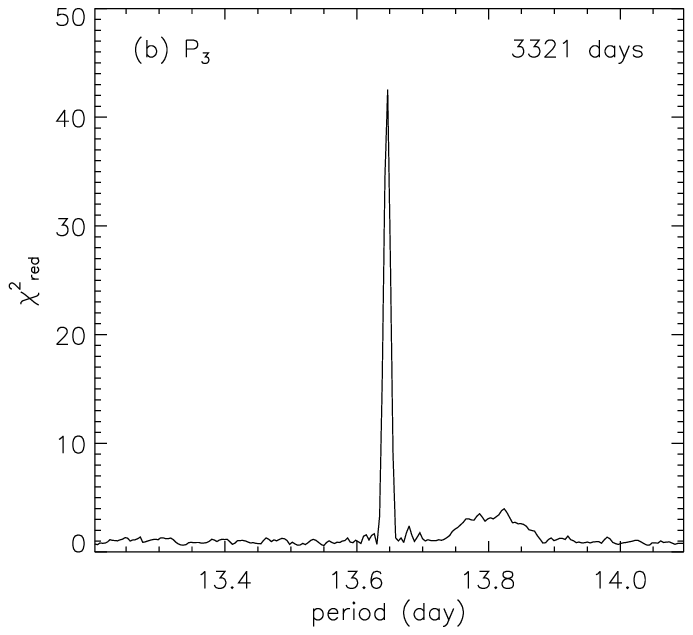}
\plotone{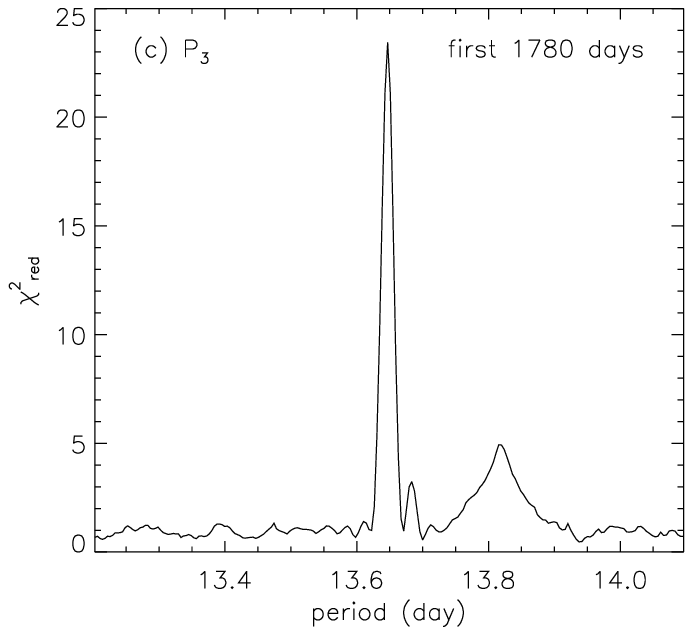}
\plotone{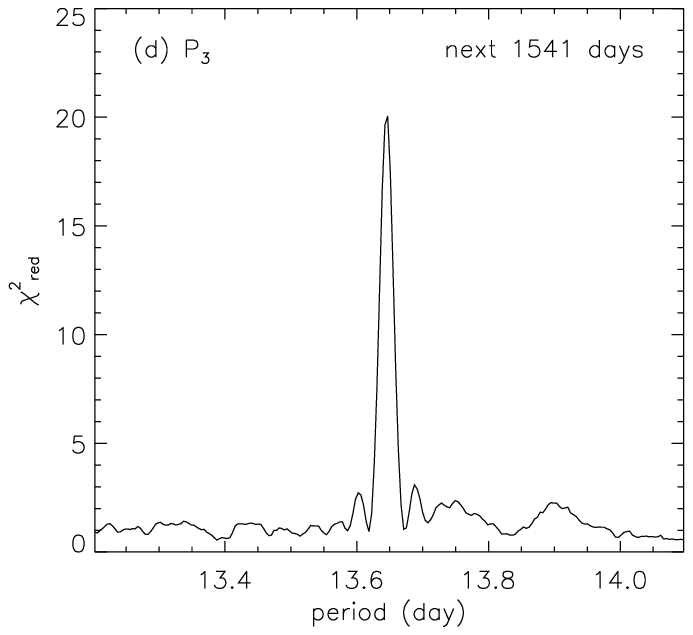}
\end{center}
\caption{The detailed view of epoch folding \redchisq from 
the \BATSE data: $P_2$ (a) and $P_3$ (b) region from all the data
(3321 days from JD 2448370.0),  $P_3$ region from the first 1780 days (c)
and from the next 1541 days (d) of the data.  }
\label{fig:efbatse}
\end{figure*}

The first two plots in Figure~\ref{fig:efbatse} show a detailed view of
the $P_2$ and $P_3$ regions ($P_i = (i+1) P_0$) from the \BATSE data.  
Figure~\ref{fig:efbatse} also shows the $P_3$
region of \redchisq from the first 1780 days and the next
1541 days of the \BATSE data.  The data were divided in this way in
order to make a direct comparison with the results in
\citet{Konig97}.  

In Figure~\ref{fig:efbatse}, one can notice that there is a bump around
13.8 days in the $P_3$ region (b), which indicates the presence of 13.8
day periodicity.  Judging from the $P_2$ region (a) and the expected
width of $P_3$, the bump is not from the side-lobe of $P_3$.  The bump
is more significant from the first 1780 days of data (c) and its peak
is at 13.82(2) days, which is consistent with the result in
\citet{Konig97}.  
But there is no clear indication of the 13.8 day periodicity
in the rest of the data (d), although the epoch-folding search shows
some features around 13.8 days.

\begin{figure*} 
\begin{center}
%\begin{tabular}{c}
%\epsfig{figure=f5a.eps,width=6.0cm,clip=} 
%\epsfig{figure=f5b.eps,width=6.0cm,clip=} \\
%\epsfig{figure=f5c.eps,width=6.0cm,clip=} 
%\epsfig{figure=f5d.eps,width=6.0cm,clip=} \\
%\end{tabular}
\epsscale{0.7}
%\epsscale{0.40}
\plotone{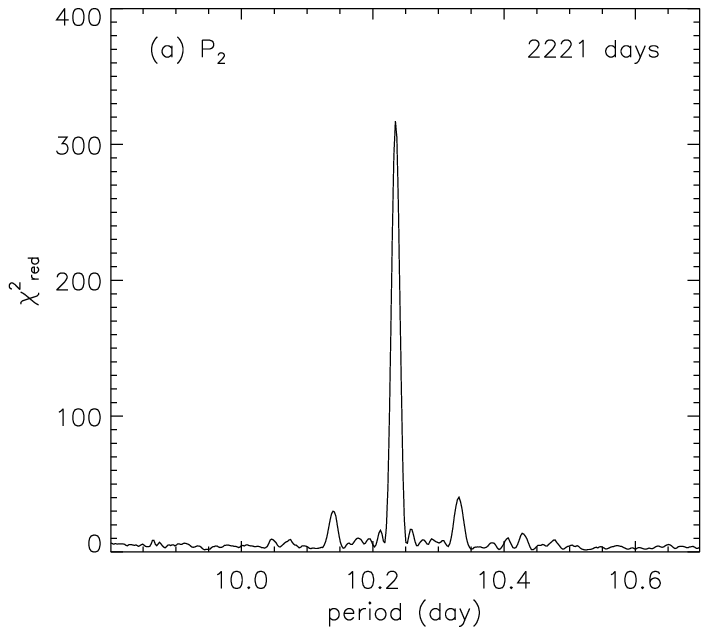}
\plotone{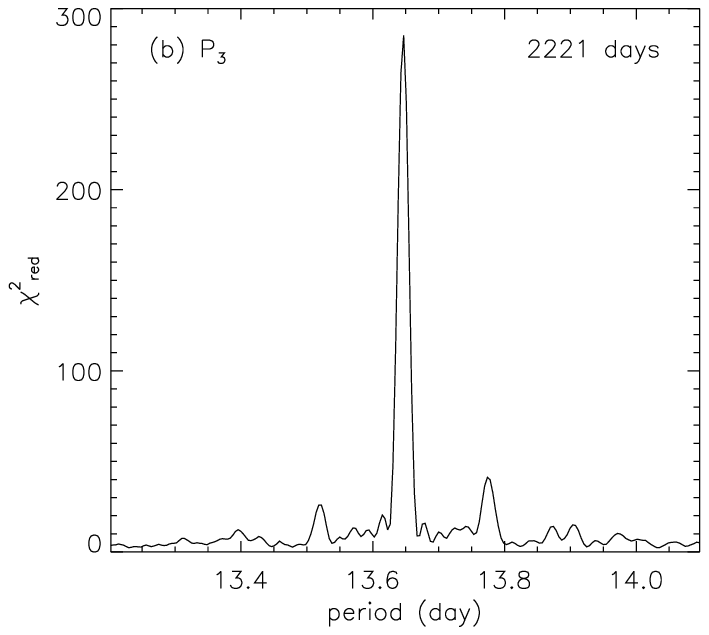}
\plotone{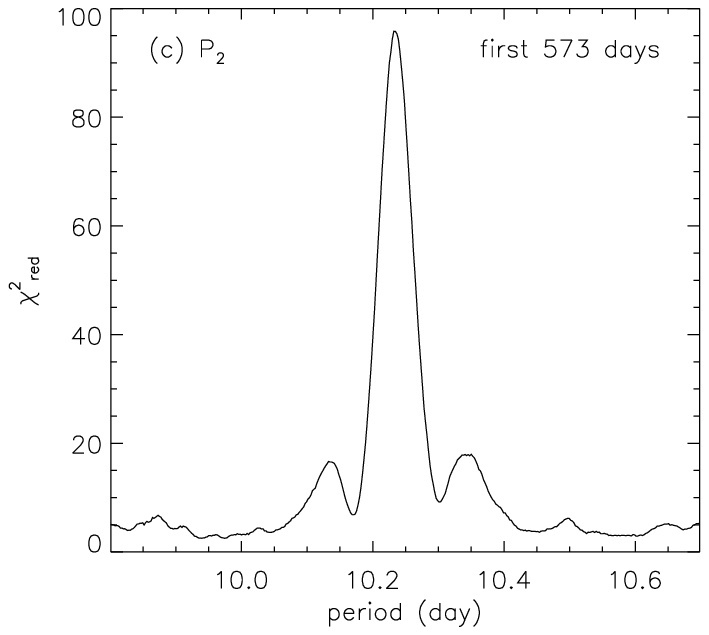}
\plotone{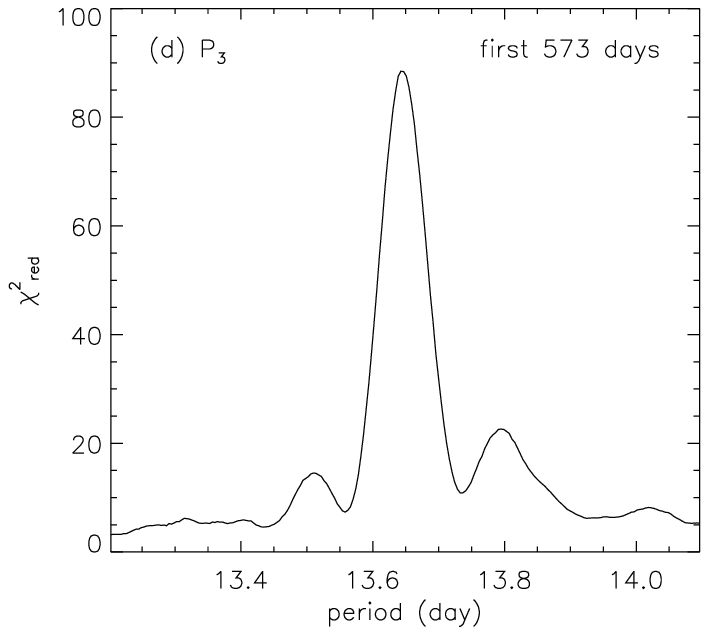}
\end{center}
\caption{The detailed view of epoch folding \redchisq
from the \ASM data:  $P_2$ (a) and $P_3$ (b) region from all the data
(2221 days from JD 2450088.6), and the same from the first 573 days (c,d)
of the data}
\label{fig:efxtet}
\end{figure*}

Figure~\ref{fig:efxtet} shows the $P_2$ and $P_3$ regions of
the epoch-folding results using the 2221 days and 
the first 573 days of the data from the \RXTE \ASM observations.

As shown in Figure~\ref{fig:efxtet} (b), there is no clear indication
of the 13.8 day peridocity from the \RXTE data.  The peak around 13.8
days appears to be a side-lobe of $P_3$, since the similar feature is
present in the $P_2$ regime (a).  The two bottom plots employ the same
573 days of the data as in \citet{Konig97}. Judging from the width of
the $P_3$ peak in the plot and its expectation ($\Delta P_3 = 4 P_0^2
/T\Ss{tot} \approx 0.08$ day where $T$\Ss{tot} = 573 days) we can
conclude that the bump around 13.8 days is largely due to the
side-lobe of $P_3$.  One could argue that the right side-lobe near
13.8 days is larger than the left side-lobe at 13.5 days in the
bottom-right plot, which may be interpreted as the presence of an
additional bump, but the similar trend is noticed in other $P_i$
regions.

We divided the data into several sets (for example, dividing into 700
days of data for each set provides 5 sets of \BATSE and  3 sets of
\RXTE data) and applied the epoch-folding technique (in the case of
\RXTE, a separate light curve is available for three energy bands). All
the results (not shown here) indicate that the 13.8 day
periodicity in the \BATSE data is significant only in the first
$\sim$ 1700 days of the data, which were taken
before the \RXTE observation on the source started.

\begin{figure*} 
\begin{center}
\epsscale{1.2}
%\epsscale{0.7}
\plotone{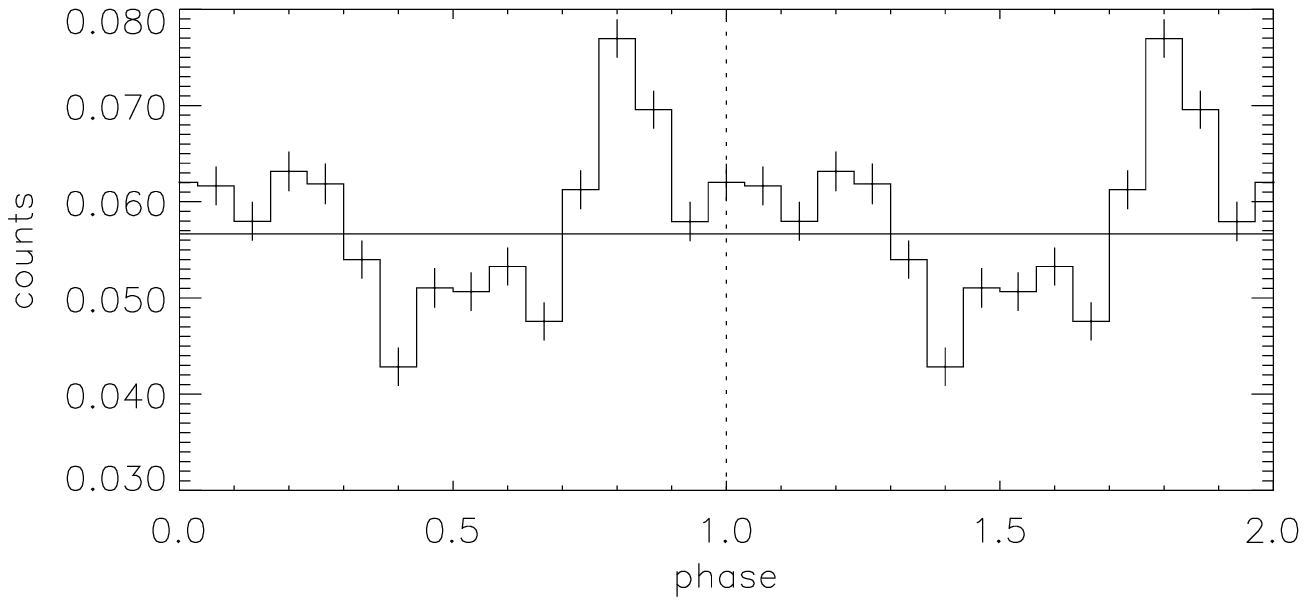} 
\plotone{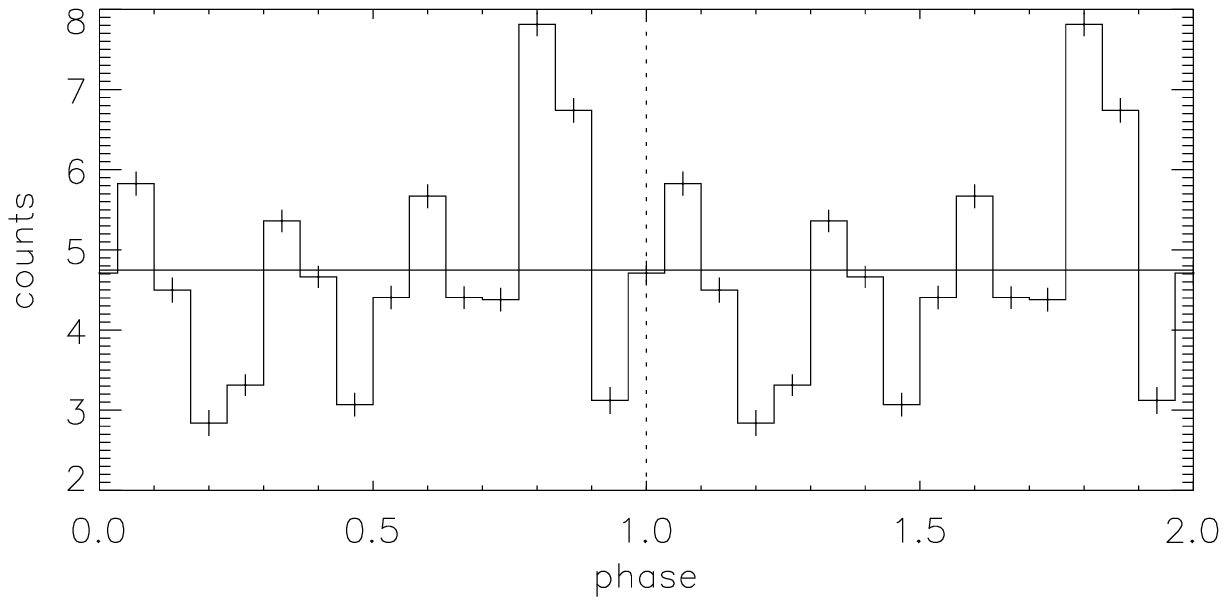}
\end{center}
\caption{Folded pulse profiles at 13.8 day; the first 1780 days
of the \CGRO \BATSE data (top), and the first 573 days of the \RXTE \ASM data (bottom), 
zero is JD 2448369.0}
\label{fig:phase}
\end{figure*}

Figure~\ref{fig:phase} shows the folded light curve at the 13.8 day
period, using the first 1780 days of the \CGRO \BATSE and the first 573
days of the \RXTE \ASM data. In the case of the \ASM data, there is a
series of small peaks caused by the proximity of $P_3$ to the 13.8 day
period for the total time span (573 days).  However, the peak at the
0.8 phase in the \ASM data is similiar to the one in the \BATSE data,
which is considered the main source of the high \redchisq value at the
13.8 day period in the epoch folding analysis.  This implies that the \ASM
data may contain a long term periodicity and its signature is too weak
to be directly detected, particularly due to the proximity of $P_3$.

In order to detect the 13.8 day periodicity unambiguously, it is
important to remove the contribution from the side lobe of $P_3$ in
calculating \redchisq for a given period.  Therefore, we applied the
epoch folding procedure only on the uneclipsed data, the orbital phase
of which falls in between 0.2 and 0.8.  If a folded phase bin at a
certain period (usually around the multiple of the orbital period)
contains no data due to exclusion of the eclipsed data, we simply
ignore the bin for calculation of the \redchisq at the period\footnote{
In the epoch folding procedure, $\chi^2=\Sigma (x_i -
\bar{x})^2/\sigma_i^2$, where $x_i$ is the counting rate in the $i$th
phase bin, $\bar{x}$ the average counting rate. Under the null
hypothesis, the standard deviations $\sigma_i$ should be defined as
$\sigma_i^2 = \sigma_{\mbox{\scriptsize tot}}^2/n_i$, where
$\sigma$\Ss{tot} is the standard deviation of the unfolded time series
and $n_i$ is the number of data points in the $i$th phase bin
\citep{Larsson96}.  Therefore, when $n_i=0$, the $i$th phase bin
contributes nothing to $\chi^2$.}.

\begin{figure*} 
\begin{center}
\epsscale{1.3}
%\epsscale{0.65}
\plotone{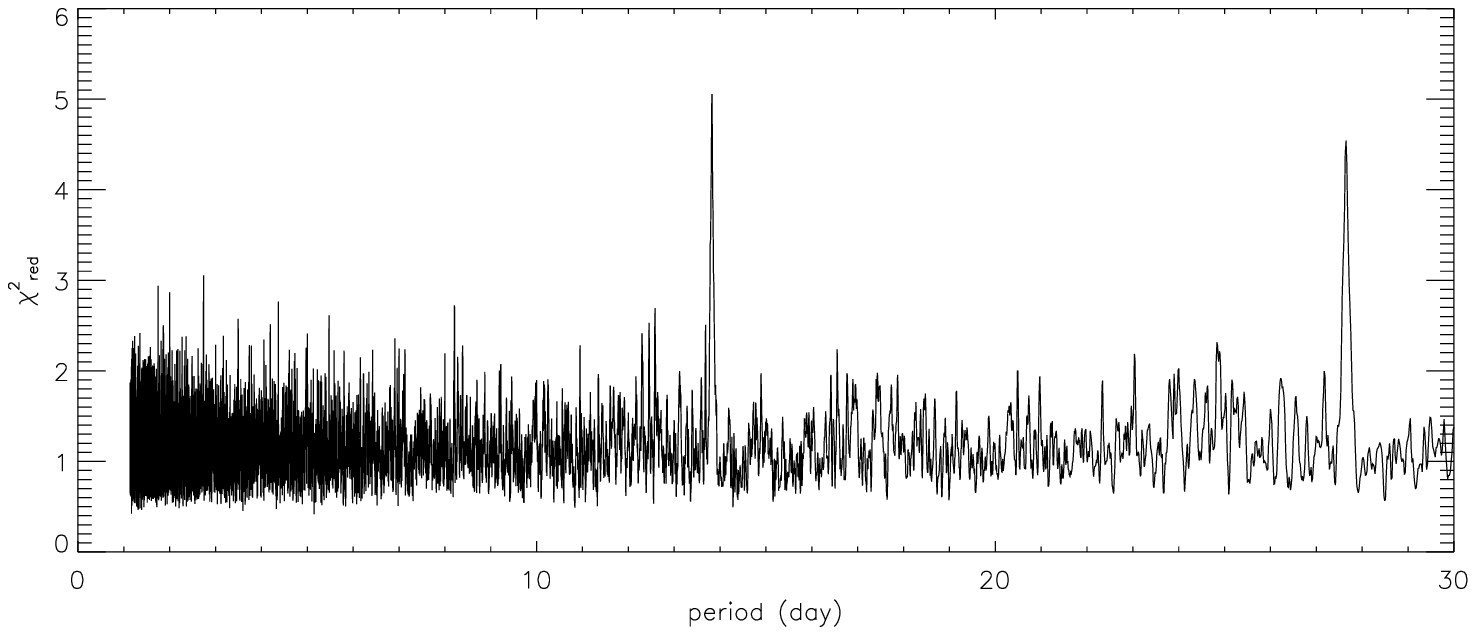}
\epsscale{0.65}
%\epsscale{0.325}
\plotone{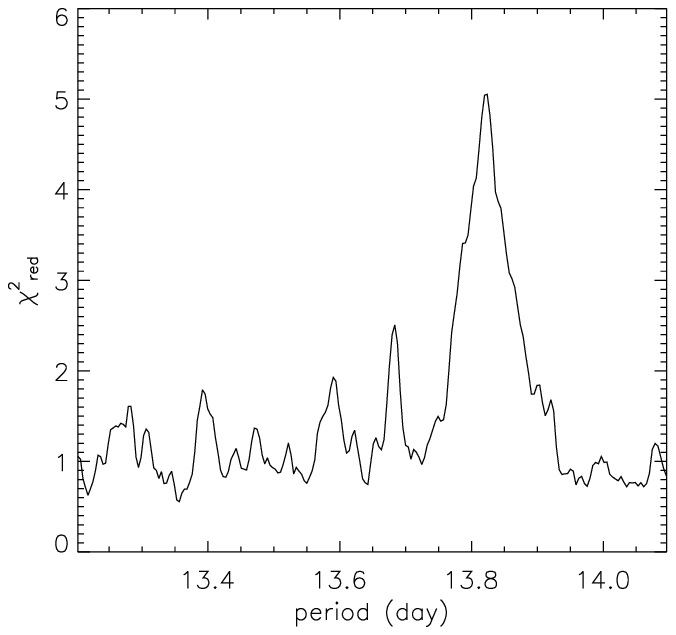}
\epsscale{1.3}
%\epsscale{0.65}
\plotone{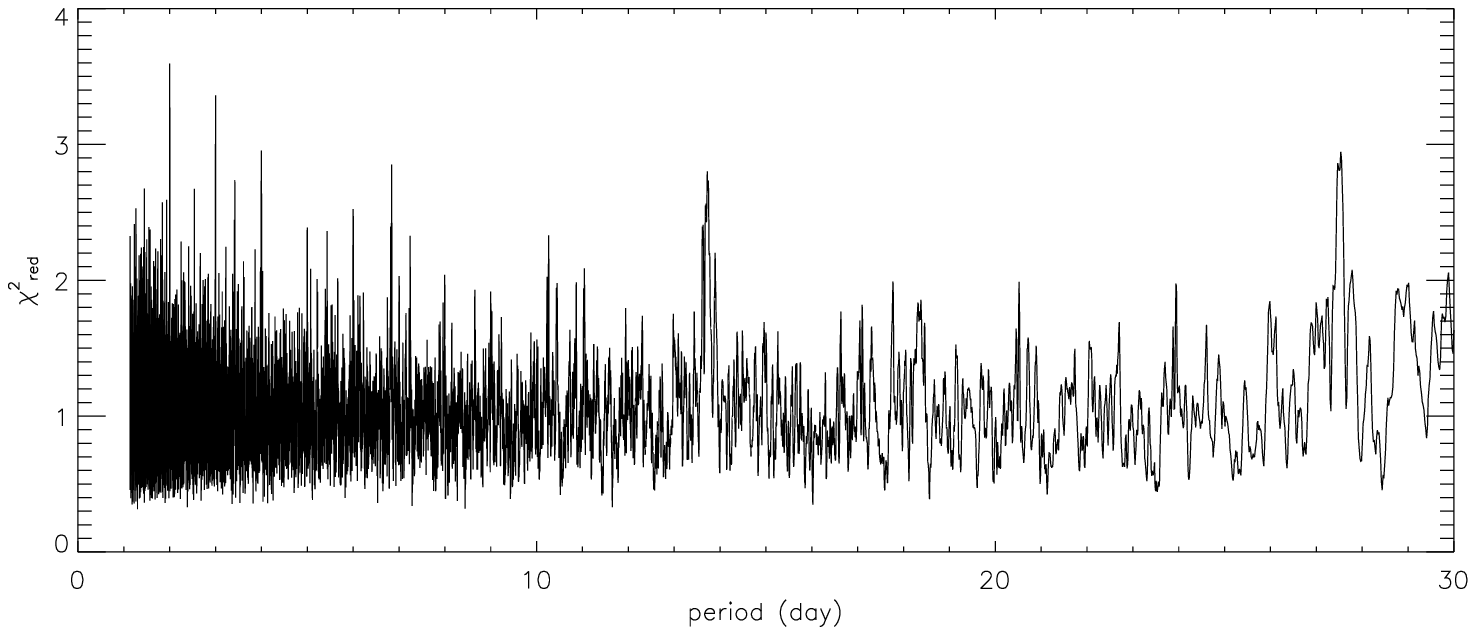}
\epsscale{0.65}
%\epsscale{0.325}
\plotone{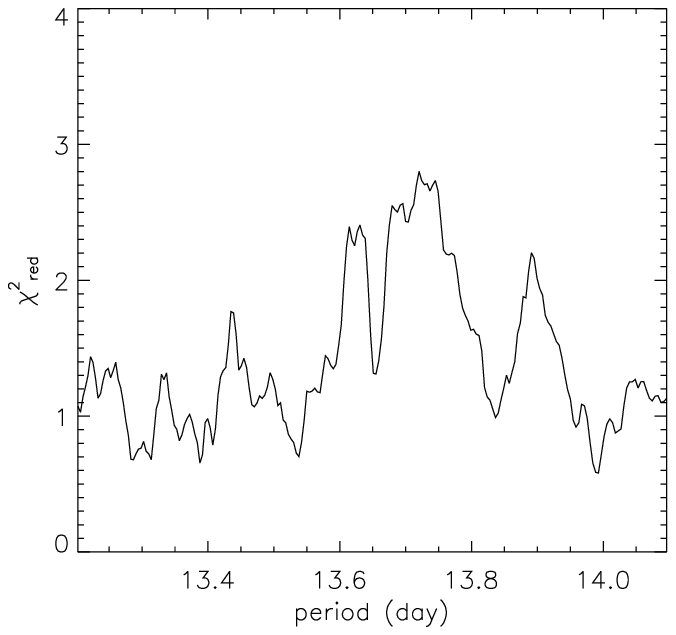}
\epsscale{1.3}
%\epsscale{0.65}
\plotone{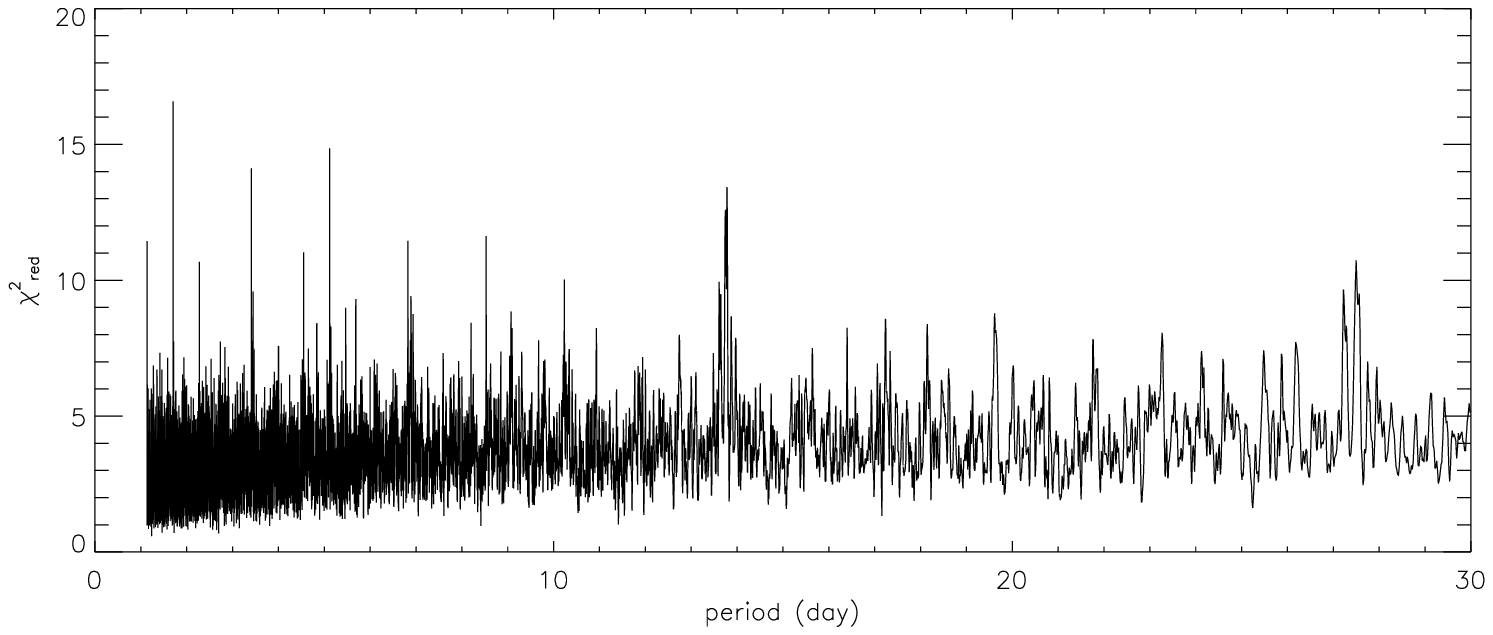}
\epsscale{0.65}
%\epsscale{0.325}
\plotone{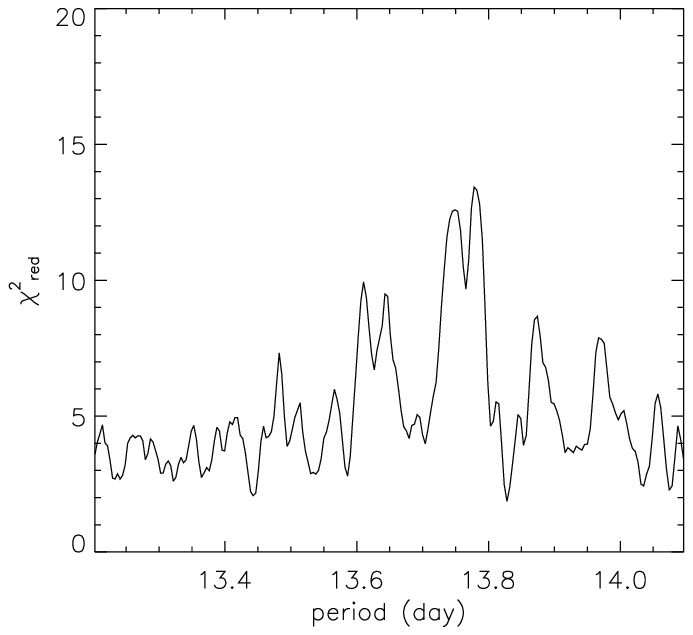}
\end{center}
\caption{The phase-averaged epoch-folding results on the uneclipsed data:
the first 1780 days of the \BATSE data (top),
the rest of the \BATSE data (middle), all the \ASM data (bottom).
The right panels are a close-up around 13.8 days for each case.
The long term periodicity and its second harmonics are clearly noticeable.}
\label{fig:noeclipse}
\end{figure*}

The result is shown in Figure~\ref{fig:noeclipse} and the estimations
of the long term periodicity are summarized in Table~\ref{tab:long}.
The figure indicates that the 13.8 day periodicity became weak but it
is clearly present in the \RXTE data and the later part of the \BATSE
data.  The peak position of the \redchisq around 13.8 day down-shifted
and the peak appears to have spread, indicating that the peak consists
of more than one periodicity (Table~\ref{tab:long}). In
Figure~\ref{fig:noeclipse}, one can also notice its second harmonics as
well as the 13.8 day periodicity. In the next section, we will discuss
the origin of the long term periodicity and its weakening.

\begin{table*}
\caption{Estimate of the long term period $P_l$ and its width $\Delta P_l$
in the epoch-folding diagram}
\label{tab:long}
\begin{tabular}{r|ccr}
\hline \hline
Data set		& $P_l$ estimate	& $\Delta P_l$ estimate/theory	& \redchisq \\
	days		& day 			& $\times 10^{-1}$ day 	  	&  	\\
\hline
\BATSE  first 1780 	& 13.82(2)		& 0.97(11)/1.07			&4.93	\\
\sS{a}first 1780 	& 13.82(2)		& 0.83(18)/1.07			&5.05	\\
\sS{a,b}next 1541 	& 13.71(3)		& 2.24(34)/1.22			&2.80	\\
	3321 		& 13.81(1)		& 1.21(36)/0.57			&3.98	\\
 \sS{a,b}3321 		& 13.81(1)		& 1.34(33)/0.57			&4.14	\\
\hline
\ASM \sS{a}2221	& 13.74(1)		& 2.60(13)/0.85			&13.5	\\
\hline
\end{tabular}
\\ \sS{a}the uneclipsed data (the orbital phase: 0.2 -- 0.8)
\\ \sS{b}exhibits multiple peaks around 13.8 days. Here we consider
them as a part of a single broad peak to estimate the overall peak
location and the width.  
\end{table*}

\section{Discussion\label{sec:dis}}

\citet{Konig97} argued that the origin of the 13.8 day periodicity is a
free precession of the neutron star arising from misalignment of the
orbital axis and the neutron star spin axis.  The analysis of the \CGRO
\BATSE and \RXTE \ASM data indicates that the 13.8 day periodicity
weakened dramatically after $\sim$ JD 2450000.  If the precession of
the neutron star caused the 13.8 day periodicity, the weakening of the
periodicity implies that the neutron star precession has been
dissipating.  Such an interpretation may not be the most favorable one
because it suggests that \4u1700 was experiencing a sudden, unusual
change of the system geometry.

In this section, we investigate accretion disk precession models in
order to explain the 13.8 day periodicity and the instability of X-ray
emission from \4u1700, without invoking a dramatic change of the system
geometry.  So far about 15 X-ray binary systems are known to show a
long term period, often called a superorbital period, in their light
curve, and in many cases, the superorbital period is successfully
explained by a precession of the accretion disk with respect to the
orbital plane \citep{Wijers99,Ogilvie01}. Although \4u1700 is a
primarily stellar wind driven system, there are several observational
results supporting a possible transient accretion disk in the system.
Therefore we first explore the clues and the characteristics of an
accretion disk in the system.  Second, we apply two different
accretion-disk precession models -- radiation driven (Ogilvie \& Dubus
2001 - hereafter OD), 
and magnetically driven \citep{Lai99} warped disk precession models 
and we study the implications.

\subsection{A long-term, transient accretion disk?}

Disk accretion is one of the most radiation-efficient accretion
processes.  When a stellar wind driven system like \4u1700 forms an
accretion disk, the disk accretion usually enhances the overall
accretion rate, leading to higher X-ray emission in general.  So a
sudden increase of X-ray flux from some systems indicates the formation
of an accretion disk in the system.  For example, GX 301--2 is a
wind-driven high-mass X-ray binary (HMXRB) with a partially underfilled
Roche lobe, similar to \4u1700.  The observed spin-up episodes in GX
301--2 are successfully explained by the formation of a transient disk
and its angular momentum exchange with the neutron star \citep{Koh97}.
Since the spin-up episodes in GX 301--2 are accompanied by enhancement
in the X-ray flux from the system, an increase in X-ray emission can be
used as indirect evidence of an accretion disk in the system.

\begin{figure*} 
\begin{center}
\epsscale{0.7}
%\epsscale{0.4}
\plotone{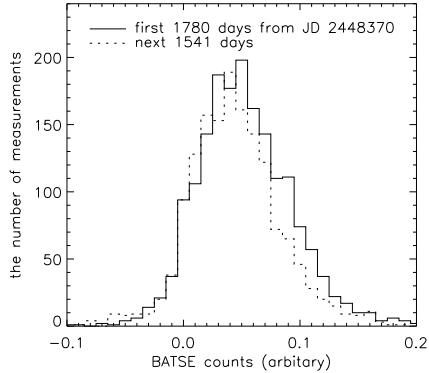}
\end{center}
\caption{The intensity distribution of the first 1780 days of the \BATSE
data (the solid line) and the next 1541 days of the data (the dotted
line): when the 13.8 day periodicity is present (the first 1780 days
of the data), there are more occurrences of high count
episodes.  The average flux in the given unit is 0.0581 for the first
1780 days of the data and 0.0498 for the next 1541 days of the days.
}
\label{fig:intensity_distribution} \end{figure*}

In the case of \4u1700, the intensity distribution of the first 1780
days of the \BATSE data shows a clear difference from that of the next
1541 days of the data (Fig.~\ref{fig:intensity_distribution}). In the
first 1780 days of the data, which exhibit the 13.8 day periodicity
rather strongly, there are relatively more occurrences of high flux
events than in the next 1541 days of the data.  The difference in the
average flux of these two data sets is 0.0083 in the given unit.
Randomly dividing the combined data into two sets of data - 1780 and
1541 day long each, we found that the probability of getting such a
difference from random fluctuations in a single distribution is
negligible ($< 10^{-8}$).

The increase in the X-ray flux may be caused by changes in the
stellar wind of the companion. However, the correlation between the
frequency of high X-ray flux episodes and the strength of the 13.8 day
periodicity indicates that both may be related to disk
accretion.  If the difference in the average flux is caused by changes
in disk accretion, the estimated fraction of the X-ray flux from disk
accretion is $\gtrsim$ 14\%.  This is a very crude estimation due to
the limited data statistics.

Other compelling evidence of disk accretion is quasi-periodic
oscillations (QPOs) of X-ray flux. It is often believed that QPOs are
caused by the material orbiting near the inner edge of the accretion
disk.  In \4u1700, QPOs were observed on several occasions during the
uneclipsed phase and they lasted $\sim$ 200 -- 1200 seconds
(\citet{Rubin96} and references therein).  In order to explain these
QPO episodes, \citet{Rubin96} proposed a beat frequency model (BFM)
mechanism operating in a transient accretion disk near the X-ray
source. In the BFM, a series of clumps in the accretion disk orbit the
neutron star at the magnetopause in Keplerian motion, and the QPOs are
a modulation at the beat frequency between the Keplerian rotation
frequency at the magnetosphere and the neutron star spin frequency.

The episodic detection of QPOs may imply a transient accretion disk
\citep{Rubin96}.  \citet{Rubin96} estimated $0.8 < \Omega < 1.0$
(Roche-lobe filling factor), which also supports the idea of accretion
through a transient disk. Taking the idea of a transient disk a step
further, we suggest that a transient disk in \4u1700 can be quite
persistent ($>$ 1000 days).

First, it should be noted that the episodic behaviour could be simply
due to the occasional formation of certain clumps in a persistent or
transient disk.  Second, if the companion O star is a
just-underfilled-Roche lobe system ($0< 1 - \Omega \ll 1.0$), the star
may frequently undergo at least partial Roche-lobe overfilling due to
various instabilities. For instance, tidal force driven disruption of
the atmosphere of the companion star can cause instability.  Naturally
the time scale for such overfilling would be related to the orbital
period -- 3.4 days.  Therefore, such instabilities provide a
semi-periodic source to feed a transient or permanent accretion
disk.

The continual Roche-lobe overfilling may result in accretion through
several successive rings rather than a continuous disk.  However, as
often assumed for many disk models, it is believed that the inward
drift rate of such a ring is much slower than the rate of viscous-driven 
ring broadening.  The material is likely to be accreted by one
continuous disk rather than multiple rings.

In fact, if the 13.8 day period is due to disk accretion, one can
loosely estimate the lifetime of the accretion disk from the
epoch-folding search results.  In the epoch-folding searches of the
first 1780 days of the \BATSE data, the width of the $\chi^2$ peak around
13.8 days indicates that the peak originates in a single periodicity.

However, in the case that a series of short-term transient disks formed
and disappeared successively in the given epoch, the long term
periodicity is likely made of multiple periods.  This results because
the long term periodicity caused by the precession of each disk could
be slightly different from one another, although it could be roughly
the same as long as each disk forms and operates in similar
conditions.  Furthermore, it is not guaranteed that the periodicity in
X-rays from each disk would be in phase unless the disk formation
procedure is periodic.  The different periodicities and their distinct
phases induced by each disk can contribute to wash out the modulation of
the folded light curve in epoch folding searches, and these may have
caused the observed weakening and spreading of the long term
periodicity in the later part of the data.

Consequently, a single, relatively strong 13.8 day periodicity in the
early \BATSE data is likely due to one or only a few accretion disks, and the
duration of the strong 13.8 day periodicity is of order the
lifetime of an accretion disk in the system.  It is not easy to
estimate the exact duration of the 13.8 day periodicity in the data.
However, judging from the significance of the $\chi^2$ over $\sim$ 300
days or 1700 days of data, we estimate that the lifetime of the
transient disk could be larger than a few 100 days.

One can also estimate the disk lifetime in relation to the time scale
$T$\Ss{visc} of viscous damping of the disk.
\begin{eqnarray*}
	T\Ss{visc} \sim 3.5\ \mbox{days}\  \alpha^{-4/5} 
		\dot{M}_{16}^{-3/10} R_{10}^{5/4} M_X^{1/4},
\end{eqnarray*} where $\alpha$ is the viscosity parameter, $M_X$ the
neutron star mass in \Ms , $\dot{M}_{16}$ the accretion rate in 
$10^{16}$ g/sec, and $R_{10}$ the radial distance from the neutron star in
$10^{10}$ cm \citep{Frank92}.

The accretion rate can be calculated from the observed flux $F_X \sim
10^{36}$  erg/sec, and $F_X$ is mostly due to stellar wind capture.
Assuming that the radiation efficiency $\epsilon$ is $\sim$ 0.1 and
that $\sim$ 14\% of $F_X$ is from disk accretion, we get $\dot{M}_{16}
\sim 0.16$.  The range of $R$ for the disk can be estimated, too.
The disc is cut off at the tidal radius $R$\Ss{tides} where the tidal
torque first changes sign, and thus the outer radius of the disk $r_o$ =
$R$\Ss{tides}. According to \citet{Wijers99}, $R$\Ss{tides} $\approx$ 0.87
$R_L$, where $R_L$ is the volume-averaged Roche radius of the accretor.
\citet{Eggleton83} calculated $R_L$ in terms of the binary separation
$r_b$ and the mass ratio $q_X$ (= $M_X$/$M_O$ = $1/q$, where $M_O$ is
the mass of the companion star).  Assuming $r_b \approx 31 \Rs $ and $q
\approx 12$  for \4u1700 \citep{Rubin96}, we get $r_o\sim 3.8 \times
10^{11}$ cm.  Finally, according to \citet{Rubin96}, $M_X$ $\approx$
2.6 \Ms.

Now we can estimate the disk lifetime from $T$\Ss{visc} at $R$ =
$r_o$.  In \4u1700, at $R$ = $r_o$, we gets $T$\Ss{visc} $\sim$ 1900 --
4600 days for a typical $\alpha$ in the range of 0.1 -- 0.3.  As
mentioned earlier, the inward drift in the disk is mainly driven by the
viscosity, and therefore, such a long viscous time scale implies that
the disk can have a long lifetime ($>$ 1000 days).

In summary, \4u1700 can have a transient disk that can last $>$ 1000
days due to frequent Roche-lobe overfilling ($\sim$ 3.4 days).  If the
precession of such an accreton disk induces the 13.8 day period in
X-ray emission, the change of the periodicity is likely related to the
stability of the precession.  In addition, in this picture of a
frequent Roche-lobe overfilled accretion disk model, instability of the
accretion disk itself can contribute to the weakening of the 13.8 day
periodicity in X-ray emission. For instance, the system may have
transitioned into a less frequent Roche-lobe overfilling state that
allows the formation of very short-term transient disks only. This is
consistent with the observed weakening and spreading of the long term
periodicity in the later part of the data.

\subsection{Radiation driven disk warping and precession}

\begin{figure*} \begin{center}
\epsscale{0.7}
%\epsscale{0.4}
\plotone{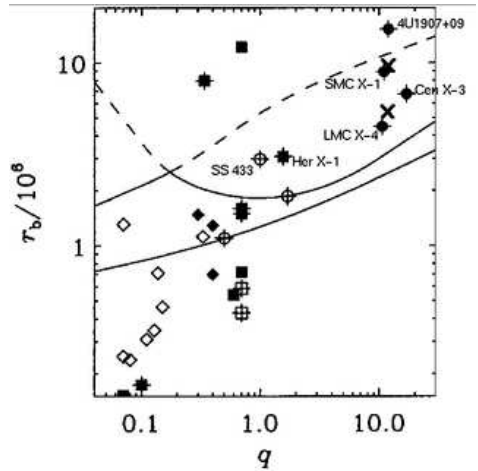}
\end{center}
\caption{Stability of X-ray binaries to radiation-driven warping
($\alpha = 0.3$ and $\epsilon = 0.1$), taken from
\citet{Ogilvie01}. The lines represent the low boundary values of
$r_b$ for the instability condition
for a given $q$, depending on the bending modes. 
Two estimates for \4u1700 in the text are marked with an X.  HMXRBs are
shown with circles, soft X-ray transients with diamonds, other low
mass X-ray binaries as squares.  Black hole candidates appear as open
symbols and neutron star primaries as solid symbols.  Crosses
represent systems for which a long term period has been reported. See
Table~1 in \citet{Ogilvie01} for the list of the binaries in this
figure.} \label{fig:stability}
\end{figure*}

In the radiation-driven, warped-disk precession model, the radiation
from the central source provides a torque to warp the disk and 
subsequently allows the precession of the disk.
The instability criterion for the radiation-driven disk warping
is 
\begin{eqnarray*}
	\frac{r_b}{GM_X/c^2} \gtrsim \frac{16 \pi^2 \eta^2}{\epsilon^2}
	\frac{r_b}{r_o},
\end{eqnarray*}
where $\eta=\nu_2/\nu_1$ is the
ratio of the two effective viscosities associated with the shear
parallel ($\nu_1$) and vertical ($\nu_2$) to the disk plane
(OD).
Under the assumptions of the model by OD, $\eta$ can be
expressed as a function of the viscosity parameter $\alpha$.
In \citet{Papaloizou77}, the ratio, $r_o/r_b$, can be given by
a function of the mass ratio $q$.  Therefore, once $\alpha$ and
$\epsilon$ are known, one can determine the instability of disk
warping from $r_b$ for a given $q$.  Figure~\ref{fig:stability},
taken from OD, shows the instability condition in
$r_b$ versus $q$ ($r_b$ is now in $GM_X/c^2$). 

In the figure, OD pointed out that systems near the instability limit
tend to show a steady precession ({\it e.g.} Her X-1, SS 433, and LMC
X-4).  The systems that lie away from the instability limit exhibit
somewhat unstable precession ({\it e.g.} Cen X-3, SMC X-1, 4U1907+09).

In the case of HMXRBs ($q$ $\sim$ 10), the instability condition
becomes $r_b/10^6 \gtrsim 3.5$ for $\alpha = 0.3$ and $\epsilon = 0.1$
(OD).  In the case of \4u1700, assuming $r_b \approx$ 31 $\Rs$ and $M_X
\approx 2.6 \Ms$ \citep{Rubin96}, $r_b/10^6$ is $\sim$ 5.4. Or if $r_b
\approx$ 36 $\Rs$ and $M_X \approx$ 1.8 $\Ms$ \citep{Heap92},
$r_b/10^6$ is $\sim$ 9.4.  While \4u1700 does satisfy the instability
condition of disk warping, it lies somewhat away from the instability
limit.  Therefore, according to OD, \4u1700 belongs to the group of
unstably precessing HMXRBs such as Cen X-3, SMC X-1, and 4U 1907+09.
Therefore, if the 13.8 day period of X-ray flux from \4u1700 is due to
a radiation driven disk precession, the instability of the periodicity
is not surprising.

It should be noted that in the above comparison with the model, many
parameters assumed for \4u1700 are not well constrained.  In
particular, $\alpha$, a critical parameter of the model,  is unknown
for the disk of \4u1700.  Thus, as in OD, we assumed $\alpha$ = 0.3,
which is a typical value for a nominal disk and this assumption allows
a direct comparison with the model described in OD.

The precession period estimate given by OD also
indirectly supports the idea of the unstably precessing disk in
\4u1700.   The estimate in OD is given only for steadily
precessing systems and the estimated period $P\Ss{long}$ depends mostly on $M_X$ and
$\dot{M}$.  For $q \ge 1$, the range of precession frequency $w_p$ is 
\begin{eqnarray*} 
	w_p & \sim &(3.2 - 7.0) \times 10^{-7} C_{\mathcal{I}}^{7/10}
		(\dot{M}/2\pi)^{3/10} (G M_X)^{-3/2} c^{5/2} \\
	 & \propto & P^{-1}_{\mbox{\scriptsize long}}
\end{eqnarray*} where $C_{\mathcal{I}}$ is a constant relating 
the second vertical moment of the density and the surface density
of the disk ($C_{\mathcal{I}}$ = 4.4 in cgs units ; refer to Eq.~9 in OD). 
The predicted precession period given for Her X-1 in OD
allows one to estimate the precession period for other steadily precessing system.
\begin{eqnarray*} 
	P\Ss{long} = (22 - 47 \ \mbox{days}) 
		\left( \frac{40}{\dot{M}_{16}} \right)^{3/10}
		\left( \frac{M_X}{1.4\Ms} \right)^{3/2}
\end{eqnarray*} 

If \4u1700 is a steadily precessing system, the estimated range of the
long term period is $\sim$ 290 -- 620 days for $M_X =$ 2.6 $\Ms$ and
$\dot{M}_{16} \sim 0.16$.  Therefore, the observed 13.8 day period is
too short for \4u1700 to be a steadily precessing system according to
the model.  However, OD also stated that the prediction is subject to
large errors due to large uncertainties in the model parameters and for
\4u1700, $\dot{M}_{16}$ is somewhat poorly constrained here.

\subsection{Magnetically driven disk warping and precession}

\citet{Lai99} introduced a magnetically driven warping/precession model
for superorbital periods in X-ray binaries. In this model, strong
magnetic fields in the magnetosphere, in particular ones penetrating
the accretion disk, exert a torque necessary for disk warping and
precession.  In magnetically driven disk warping an inner portion
of the disk will be warped and as a result the disk will precess.

In the case of \4u1700, persistent pulsation has not been observed.
Therefore, the existence of a strong magnetic field in the system is
not confirmed.  However, the hard X-ray spectrum 
implies that the system may have strong magnetic fields. For example,
a hard tail in the hard X-ray spectrum is the characteristic of
a massive binary system with an X-ray pulsar \citep{Laurent92}, and the
{\it BeppoSAX} observation showed some hints of a broad cyclotron
absorption feature \citep{Reynolds99}.

The instability condition of magnetically driven warping is that
the magnetic warping gross rate $\Gamma_w$ should be larger than the
viscous damping rate \citep{Lai99}, or
\begin{eqnarray*}
	\Gamma_w^{-1} \lesssim r^2/\nu_2 = T\Ss{visc} (r),
\end{eqnarray*} where $r \lesssim r_m$ and
$r_m$ is the radius of the magnetosphere boundary, and
\begin{eqnarray*}
	r_m &= & (3.4 \times 10^8 \mbox{cm})\  \eta\ 
		\mu_{30}^{4/7} M_{1.4}^{-1/7} \dot{M}_{17}^{-2/7},
\end{eqnarray*} where $\eta$ is a constant of order unity  and  $\mu_{30}$
the magnetic dipole moment of the neutron star in $10^{30}$ G cm$^3$
\citep{Lai99}.

For example, in the case of \4u1700, $T$\Ss{visc} $\sim$ 6 day at $r$
$\sim$ 0.01 $r_o$.  Since $\Gamma_w^{-1} \propto r^{49/10}$ and
$T\Ss{visc} \propto r^{5/4}$, if $\Gamma_w^{-1} \lesssim$ 6 day at $r$
= 0.01 $r_o$, $\Gamma_w^{-1} < T\Ss{visc} (r)$ at $r< 0.01$~$r_o$.
Then, the magnetic fields can warp the disk at $r < 0.01$~$r_o$.  For
\4u1700, assuming $M_{1.4}$ = 1.86 (= $M_X$/1.4$\Ms$), 
$\dot{M}_{17} \sim 0.016$, and
$\eta \sim 1$,  we get $r_m \sim$ (1.01$\times$$10^{9}$~cm) $\mu^{4/7}_{30}$.
Therefore, in order to have magnetically driven warping at $r <
0.01$~$r_o$, $\mu_{30} > 10$ from $r_m > 0.01$~$r_o$.

\citet{Lai99} provides $\Gamma_w$ and the magnetically driven precession
frequency $\nu_{\mbox{\scriptsize prec}}$ in terms of
various parameters such as the magnetic dipole moment.
Because of the lack of the confirmed persistent pulsation from
\4u1700, it is difficult to estimate basic information related to
the magnetic field of the system such as the magnetic dipole moment.
Thus, it is even more difficult to estimate $\Gamma_w$ and
$\nu_{\mbox{\scriptsize prec}}$.
However, we can get a magnetic-field independent relation from
the ratio of $\nu_{\mbox{\scriptsize prec}}$ and
$\Gamma_w$ (Eq 6.3 and 6.13 in \citet{Lai99}).
\begin{eqnarray*}
	\frac{\nu_{\mbox{\scriptsize prec}}}{\Gamma_w}
		 = & -\left(\displaystyle 
		\frac{5.3 \mbox{\ day}}{26 \mbox{\ day}}\right) 
		\zeta (\cos \beta \tan^2 \theta) D^{-1},
\end{eqnarray*} where $\beta$ the angle
between the neutron star spin axis and the disk rotation axis, and
$\theta$ the angle between the neutron star spin axis and its magnetic
moment, 
$D = $ max($\sqrt{r^2/r_m^2-1},\sqrt{2H/r_m}$) 
($H$ is the half-thickness of the disk and 
$\zeta$ specifies the azimuthal pitch of
the field line (in general $|\zeta| \lesssim 1$).

For a precession of the whole disk, a modal analysis is required for a
proper evaluation of $\cos \beta$.  In general, $\beta$ will alternate
between its maximum and minumum along the azimuthal direction.
Thus, azimuthally averaged $\overline{\cos \beta} \lesssim 1$.
Assuming $\zeta \sim 1 $, $D \sim 0.2$ \citep{Lai99}, $\cos \beta \sim 1$ and
$\nu_{\mbox{\scriptsize prec}} = 1/13.8 $ days, 
the above relation becomes
\begin{eqnarray*}
	\Gamma_w^{-1}  \sim (14.1 \mbox{\ day}) \tan^2 \theta.
\end{eqnarray*} 
For $\alpha = 0.3$, $\dot{M}_{16} = 0.16$, and $M_X = 2.6 \Ms$,
\begin{eqnarray*}
	T\Ss{visc}  \sim (20.2 \mbox{\ day})\  r_m^{5/4} \sim 
		(1.15 \mbox{\ day})\ \mu^{5/7}_{30},
\end{eqnarray*}
at $r=r_m$, where $r$  and $r_m$ are in  $10^{10}$ cm.
Therefore, if a magnetically field driven disk
precession caused the 13.8 day period in the X-ray emission from
\4u1700,
\begin{eqnarray*}
	\tan^2 \theta \lesssim 1.43\  r_m^{5/4} \sim 0.0815\ \mu^{5/7}_{30} .
\end{eqnarray*}
This result implies that the weaker the field is, the more likely
aligned the neutron star spin axis and the magnetic dipole axis are.
For example, if \4u1700 develops magnetically driven disk warping at $r
<$ 0.01 $r_o$ ($\Gamma_w^{-1} < 6$ day), the above relation reduces to
$\theta \lesssim 33$\sS{o} and $\mu_{30} \sim 10$.  Or in the case that
relatively weak magnetic fields are present, the warping would occur in a
smaller portion of the disk.  So, if the warping occurs at
$r<$ 0.002 $ r_o$, $\Gamma_w^{-1} \lesssim 0.81$ day, $\theta
\lesssim 13$\sS{o}, and $\mu_{30} \sim 0.61$.
If there is indeed such an alignment, it explains the lack of the
observed persisent pulsation from \4u1700.  

In disk precession models, the central source will be succesively seen
in `on' and `off' states due to the varying obscuration by the disk.
As a result, a large modulation is expected in the detected light
curve, as seen in the well-established disk-precession sources like
Her X-1, LMC X-4, and SMC X-1. However, \4u1700 exhibits a somewhat
small modulation in the folded light curve at the observed long term
period (Figure~\ref{fig:phase}).  The small modulation can be explained
consistently in the disk precession models.  First, there can be some
extended emission regions such as disk corona sources \citep{Ogilvie01}.
Second, in \4u1700, it is possible that due to the inherent instability
of disk precession (or the disk itself), for a certain fraction of
precession cycles the central source is not completely hidden by the
disk even at the maximum obscuration.  An alternative explanation is the
variable-accretion rate model, which is favored for the observed long term
period in LMC X-3 \citep{Wilms01}.  Further studies such as the spectral
variation analysis are required to assess whether such a model is compatible
with \4u1700. 

\section{Conclusion}

We studied the long term -- 13.8 day periodicity in X-ray flux from
\4u1700.  The epoch folding analysis indicates that the periodicity was
relatively strong and consisted of a single period in the early data of
the \BATSE observation.  The long periodicity feature weakened
and spread over multiple periods after $\sim$ JD 2450000 when \RXTE
started its observation.

We investigated the origin of the 13.8 day period in X-rays using disk
precession models.  Disk precession models are a natural way to explain
the 13.8 day period and its instability.  In particular, the
instability of the 13.8 day period is consistent with the
radiation-driven warped-disk precession model.  Alternatively, the
magnetically-driven warped-disk precession model can provide a
constraint on the possible alignment of neutron star spin axis and the
magnetic dipole axis, which can explain the lack of the observed
pulsation from the system.

\section{Acknowledgement}
The authors wish to thank the anonymous referee for useful discussions and
suggestions.

\end{document}